\documentclass[conference,10pt]{IEEEtran}

\usepackage{amssymb,amsmath,amsfonts,bm,epsfig,graphicx,theorem,latexsym}
\usepackage{rotating,setspace,latexsym,epsf,color,subfigure}
\usepackage{cite,authblk}
\newtheorem{remark}{Remark}

\def \n2{{N_0 \over 2}}

\def \h5{\hspace{0.5in}}

\begin{document}
\IEEEoverridecommandlockouts
\pagestyle{empty}

\title{Relative Age of Information: A New Metric for Status Update Systems}

\author{Peng Zou \qquad Omur Ozel \qquad Suresh Subramaniam \\
\normalsize Department of Electrical and Computer Engineering \\
\normalsize George Washington University, Washington, DC 20052 USA \\
\normalsize {\it pzou94, ozel, suresh@gwu.edu} }

\maketitle 

\begin{abstract}
  
In this paper, we introduce a new data freshness metric, {\em relative Age of Information (rAoI)}, and examine it in a single server system with various packet management schemes. The (classical) AoI metric was introduced to measure the staleness of status updates at the receiving end with respect to their generation at the source. This metric addresses systems where the timings of update generation at the source are absolute and can be designed separately or jointly with the transmission schedules. In many decentralized applications, transmission schedules are blind to update generation timing, and the transmitter can know the timing of an update packet only after it arrives. As such, an update becomes stale after a new one arrives. The rAoI metric measures how fresh the data is at the receiver with respect to the data at the transmitter. It introduces a particularly explicit dependence on the arrival process in the evaluation of age. We investigate several queuing disciplines and provide closed form expressions for rAoI and numerical comparisons.

\end{abstract}

%\newpage
\pagestyle{plain}
\setcounter{page}{1}
\pagenumbering{arabic}

\section{Introduction}

The timeliness of the available information arriving to or departing from interested nodes is a critical parameter in the operation of various modern communication network applications in the Internet-of-Things (IoT). Examples include a scheduler that uses time-sensitive state information from surrounding nodes, a cognitive mobile access point that utilizes channel state information for efficient transmissions in fading scenarios, and vehicular communication nodes that determine the routes of remotely controlled vehicles. Age of Information (AoI) metric, and more generally its moments and functions, have been used to measure the freshness of available information at a receiving node in such status update systems. AoI provides suitable frameworks to analyze the timeliness of information in such applications; see \cite{kaul2012real, costa2016age, inoue2018general, najm2017status, kam2018age, hsu2017age, jiang2018decentralized, yates2018status, 2018information, yates2015lazy, wu2017optimal_ieee, bedewy2017age, arafa2017age, kosta2017age2, zhong2018two, kam2018towards}.

\begin{figure}[!t]
\centering{
\hspace{-0.2cm} 
\includegraphics[totalheight=0.11\textheight]{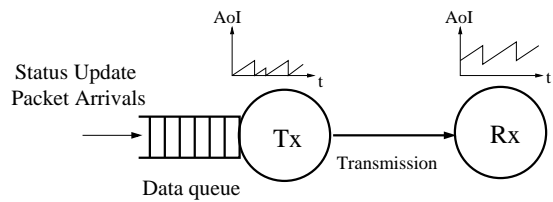}}
\caption{\sl System model with status update packets arriving to a single server queue with AoI evolutions at the transmitter and receiver sides.}\vspace{-0.1in}
\label{fig:1} 
\end{figure}

We consider  a point-to-point status update system, as shown in Fig. \ref{fig:1}, in which status update packets are generated by a source and immediately arrive to a queue to be transmitted to a receiver. \textit{The classical definition of status update age, i.e., AoI} is the time elapsed since the last received update was generated. In this paper, we introduce a new data freshness metric for a point-to-point system which we term {\em relative Age of Information (rAoI)}. The rAoI metric is simply the AoI observed at the receiver Rx relative to the AoI at the transmitter Tx. In other words, the latest update packet generated at the source is considered to be fresh, and the rAoI measures how far behind the update at the receiver is with respect to the fresh update at the source.  
%In a variety of decentralized applications, it is either costly or infeasible to track the timing of packets starting from its generation time and a reference time is needed to count the age of the packet. rAoI is the natural extension of AoI to capture the need for a frame of reference. 
In many sensor network applications, update packet generation is independent of and oblivious to the transmission process, and the source has no knowledge of the transmitter state. In this case, the packet generation and communication processes are naturally decoupled. An update becomes stale only after a new update is generated, and arrival events directly impact the evolution of age at the receiving end. Our rAoI metric captures this dependence. 

In the recent literature, there have been attempts to define metrics that are related to AoI and we bring references \cite{kosta2017age2, zhong2018two} to attention as the papers that are closest to the rAoI metric. In \cite{kosta2017age2}, non-linear age and value of information are introduced as new types of AoI metric. These metrics allow non-linear growth of age with time and the drop in the age at the event of completion of service is also addressed. In \cite{zhong2018two}, a new metric for freshness of cached information is introduced, namely age of synchronization, and it is compared with AoI in the context of cache freshness. Age of synchronization measures the time difference between the current time and the last time the most recent generated update is fetched. Earlier papers, such as \cite{costa2016age}, consider packet management and provide insights into reducing AoI by discarding packets with longer age in the queuing phase. Still, none of the earlier works exclusively examine the critical role of the arrival process in designing AoI metrics and it is one of our goals with the rAoI metric to capture this phenomenon.

In this paper, we investigate the moments of rAoI where stochastically generated updates arrive according to a Poisson process and the time it takes for a packet to be transmitted has a general distribution. We consider M/GI/1 with preemption, M/GI/1/1 and M/GI/1/$2^*$ queuing disciplines compatible with Kendall notation (see, e.g., \cite{costa2016age,inoue2018general}). Common to these schemes is that the updates arrive to the transmitter according to a Poisson process, and the time it takes for a packet to be transmitted is a random variable that has a general distribution, is independent over time and independent of other events in the system. Additionally, at most one buffer is available and a packet under service may or may not be preempted when a new update arrives. We perform stationary distribution analysis to obtain expressions for moments of rAoI and provide numerical results for various system parameters. 

\section{The Model and rAoI Metric}
\label{sec:Model}

As shown in Fig. \ref{fig:1}, we consider a point-to-point communication system with a single transmitter (Tx) and a single receiver (Rx). The status update packets arrive at the transmitter according to a Poisson process with arrival rate $\lambda$. The transmitter node transmits the status update packets one at a time. The time for a packet to be served has a general distribution $f_S(s)$, $s \geq 0$, independent of other system variables and independent over time. Corresponding to the general distribution, we have $MGF^{(S)}_{\gamma}$, the moment generating function of the service distribution at $-\gamma$ for $\gamma \geq 0$:
\begin{align}
MGF^{(S)}_{\gamma} \triangleq \mathbb{E}[e^{-\gamma S}]
\end{align}

We let $t_i$ denote the time stamp of the event that packet $i$ enters the queue, and $t_i'$ the time stamp of the event that the packet $i$ (if selected for service) is delivered to the receiver. We also denote the inter-arrival time between $t_i$ and $t_{i+1}$ as $X_i$, which is an independent exponentially distributed random process. The instantaneous Age of Information (AoI) at the receiver (transmitter) is the difference of the current time and the time stamp of the latest delivered packet at the receiver (latest arriving packet at the transmitter): 
\begin{align}
\Delta_R(t)=t-u_R(t)\\
\Delta_T(t)=t-u_T(t)
\end{align}
where $u_R(t)$ and $u_T(t)$ are the time stamps of the latest received packet by the receiver and the latest arriving packet at the transmitter, respectively, at time $t$. We can express $u_T(t)=\max\{t_i: \ t_i \leq t\}$ and $u_R(t)=t_{i^*}$ where $i^*=\max\{t_i: \ t_i' \leq t\}$. The classical AoI is $\Delta_R(t)$. The relative Age of Information (rAoI) at time $t$ is
\begin{align}
\Gamma(t)=\Delta_R(t) - \Delta_T(t)
\end{align}
$\Gamma(t)$ measures the freshness of the update available at the receiver with respect to the transmitter. As packet generation is oblivious to the transmitter state, it is naturally decoupled from the communication process. $\Gamma(t)$ measures the transmitter's performance in enabling the receiver obtain the most recent update in a timely manner. 

\begin{figure}[!t]
\centering{
\hspace{-0.2cm} 
\includegraphics[totalheight=0.2\textheight]{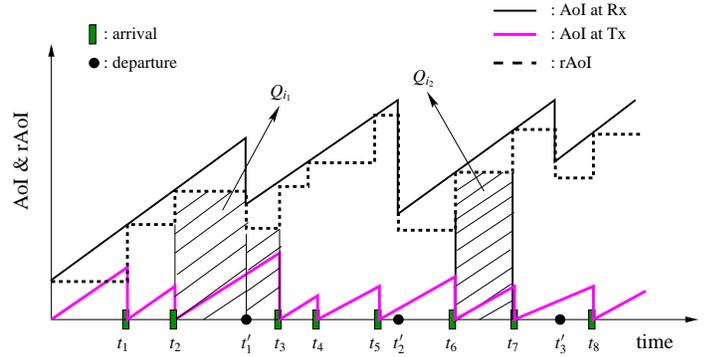}}
\caption{\sl Evolutions of AoI at the receiver ($\Delta_R$), transmitter ($\Delta_T$) and rAoI ($\Gamma$) for a FCFS queue. Note that rAoI samples $\Delta_R$ at the instants of arrivals.}\vspace{-0.2in}
\label{fig:2} 
\end{figure}

We illustrate the evolution of rAoI in Fig. \ref{fig:2}. $\Delta_T(t)$ is represented as the lower sawtooth curve that increases linearly with time and drops to zero at each arrival instant. $\Delta_R(t)$ is the uppermost curve that increases linearly in between service completion instants. The difference of these two curves represents the evolution of $\Gamma(t)$. The dotted curve in Fig. \ref{fig:2} shows clearly that the $\Gamma(t)$ curve \textit{samples} the classical AoI $\Delta_R(t)$ at each arrival instant. Then, depending on whether there is a service completion in the next inter-arrival interval, $\Gamma(t)$ either remains constant or drops to a certain value according to the state of the system. We will make use of this fact in our evaluations coming up in the next section. Note that the dependence on the arrival process is reminiscent of the AoS metric in \cite{zhong2018two}. However, different from AoS, rAoI makes a jump each time an arrival occurs, indicating that the receiver's update is behind the freshest update at the transmitter.

Customary to the practice of evaluating the expected value of AoI metrics, we define the following useful quantity:
\begin{align}
Q_i \triangleq \int_{t_i}^{t_{i+1}}\Gamma^k(t)dt
\end{align}
where $k$ is the order of the moment of rAoI we wish to calculate. In Fig. \ref{fig:2}, $Q_i$ is illustrated for $k=1$. $Q_i$ is the area under the $\Gamma^k(t)$ function between two packet arrivals in the form of sum of the areas under multiple rectangles. Due to the ergodicity of the system, the $k$th moment of rAoI is
\begin{align}
\mathbb{E}[\Gamma^k]=\lambda\lim_{N \rightarrow \infty} \frac{1}{N}\sum_{i=1}^N Q_i = \lambda \mathbb{E}[Q_i]
\end{align}
From the evolution of rAoI in Fig. \ref{fig:2} and due to the PASTA property, we make the following observation: 
\begin{remark}\label{re:2}
Average $\Gamma(t)$ at a time right after an arrival occurs is equal to average $\Delta_R$. 
\end{remark}

\section{Evaluating rAoI for Queuing Disciplines}
\label{sec:eval}

In this section, we explore several queuing disciplines that have been addressed in the literature. We obtain general expressions for $\mathbb{E}[\Gamma^k]$ in terms of known expressions for $\mathbb{E}[\Delta_R^k]$ for general service distributions, and then evaluate specifically for first and second moments under exponential and deterministic service times. Before we start, we first note the following remark.
\begin{remark}\label{re:1}
The first moment of $\Gamma$ is simply equal to
\begin{align}
\mathbb{E}[\Gamma]= \mathbb{E}[\Delta_R]-\mathbb{E}[\Delta_T]=\mathbb{E}[\Delta_R]-\frac{1}{\lambda}
\end{align} 
Hence, $\lim_{\lambda \rightarrow \infty}\mathbb{E}[\Gamma] = \lim_{\lambda \rightarrow \infty}\mathbb{E}[\Delta_R]$.
\end{remark}

\begin{figure}[!t]
\centering{
\hspace{-0.2cm} 
\includegraphics[totalheight=0.17\textheight]{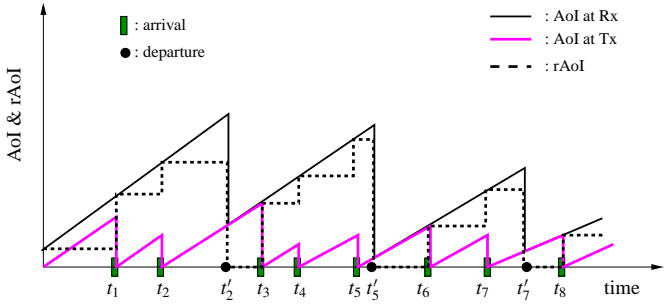}}
\caption{\sl Evolutions of $\Delta_R$, $\Delta_T$, and $\Gamma$ for M/GI/1 with preemption.}\vspace{-0.2in}
\label{fig:3} 
\end{figure}

\subsection{M/GI/1 with Preemption}

In this scheme, all incoming packets are given service right away and any packet in service is discarded. We provide a sample path of $\Delta_R$, $\Delta_T$, and $\Gamma$ in Fig. \ref{fig:3}. This is the same scenario as in Fig. \ref{fig:2} specialized to preemption. In here, at time $t_2$, the packet 1 is dropped from service and packet 2 is taken to service; its service ends at $t_2'$. At time $t_3$, packet 3 finds the queue idle and is taken to service; at time 4, packet 3 is discarded while packet $4$ is taken to service and at time $t_5$, packet 4 is discarded. Let $S_i$ denote the service time for the incoming packet. Recall that $X_i$ is the time for the next arrival and it is independent of $S_i$. 
\begin{equation}
Q_i = \left \{  
\begin{array}{cc}
 \left(\Delta_R(t_i)\right)^kX_i &  \textrm{if $X_i<S_i$}   \\
 \left(\Delta_R(t_i)\right)^kS_i  &     \textrm{if $X_i \geq S_i$}
\end{array}
\right.
\label{eq:preemption}
\end{equation}
where $\Delta_R(t_i)$ is the sample of the classical AoI at $t_i$. $X_i$ and $S_i$ are both independent of $\Delta_R(t_i)$ since AoI is determined by earlier events in the system. Observe that $\Gamma(t)$ drops to zero in the interval $[t_i+ S_i, t_i+X_i]$ if $ X_i \geq S_i$. We then have 
\begin{align}\nonumber
\mathbb{E}[Q_i]&=\mathbb{E}[(\Delta_R(t_i))^k]\int_0^\infty x P(S>x) f_X(x)dx \\ &\quad +\mathbb{E}[(\Delta_R(t_i))^k] \int_0^\infty s P(X>s) f_S(s) ds
\end{align}
Additionally, we can replace $\mathbb{E}[(\Delta_R(t_i))^k]$ with $\mathbb{E}[(\Delta_R)^k]$ due to Remark \ref{re:2}. We have the following expression:
\begin{align}\label{eq:finexp}
\mathbb{E}[Q]=\mathbb{E}[(\Delta_R)^k]\left(\frac{1}{\lambda}-\frac{1}{\lambda}MGF_{\lambda}^{(S)}\right)
\end{align}
and $\mathbb{E}[\Gamma^k]=\lambda\mathbb{E}[Q]=\mathbb{E}[(\Delta_R)^k](1-MGF_{\lambda}^{(S)})$. We can use the $k$th moment expressions for $\Delta_R$ from \cite{costa2016age,inoue2018general}.

\subsubsection{Exponential Service}

Let us consider $f_S(s)=\mu e^{-\mu s}$, $s \geq 0$. In this case, we have a memoryless service distribution and many expressions simplify. First, we note the moment generating function for $S$ is:
\begin{align}\label{eq:mgf_exp}
MGF^{(S)}_{\gamma}=\frac{\mu}{\gamma + \mu}
\end{align}
Due to \cite[Section A.2]{inoue2018general}, we have the following for classical AoI:
\begin{align}
\mathbb{E}[\Delta_R]=\frac{\lambda+\mu}{\lambda\mu}, \ \mathbb{E}[\Delta_R^2]=\frac{2(\lambda^2+\lambda\mu+\mu^2)}{\lambda^2\mu^2}
\end{align}
Then, the first moment of rAoI is $\mathbb{E}[\Gamma] =\frac{\lambda+\mu}{\lambda\mu} - \frac{1}{\lambda} = \frac{1}{\mu}$ and we express the second moment of rAoI as:
\begin{align}
\mathbb{E}[\Gamma^2]=\frac{2(\lambda^2+\lambda\mu+\mu^2)}{\lambda\mu^2(\lambda+\mu)}
\end{align}

\subsubsection{Deterministic Service}

In this case, we set $S=\frac{1}{\mu}$ with probability one for a deterministic variable $\mu$. We have the following closed form expression:
\begin{align}\label{eq:mgf_det}
MGF^{(S)}_{\gamma}=e^{-\frac{\gamma}{\mu}}
\end{align}
From \cite[Section A.2]{inoue2018general}, we have
\begin{align}
\mathbb{E}[\Delta_R]=\frac{e^{\frac{\lambda}{\mu}}}{\lambda}, \ \mathbb{E}[\Delta_R^2]=\frac{2(\mu e^{\frac{\lambda}{\mu}}-\lambda)e^{\frac{\lambda}{\mu}}}{\lambda^2\mu}
\end{align}
Then, we have the first moment as $\mathbb{E}[\Gamma] = \frac{e^{\frac{\lambda}{\mu}}-1}{\lambda}$ and the second moment of rAoI is:
\begin{align}
\mathbb{E}[\Gamma^2]=\frac{2(\mu e^{\frac{\lambda}{\mu}}-\lambda)(e^{\frac{\lambda}{\mu}}-1)}{\lambda^2\mu}
\end{align}

\subsection{M/GI/1/1}

\begin{figure}[!t]
\centering{
\hspace{-0.2cm} 
\includegraphics[totalheight=0.17\textheight]{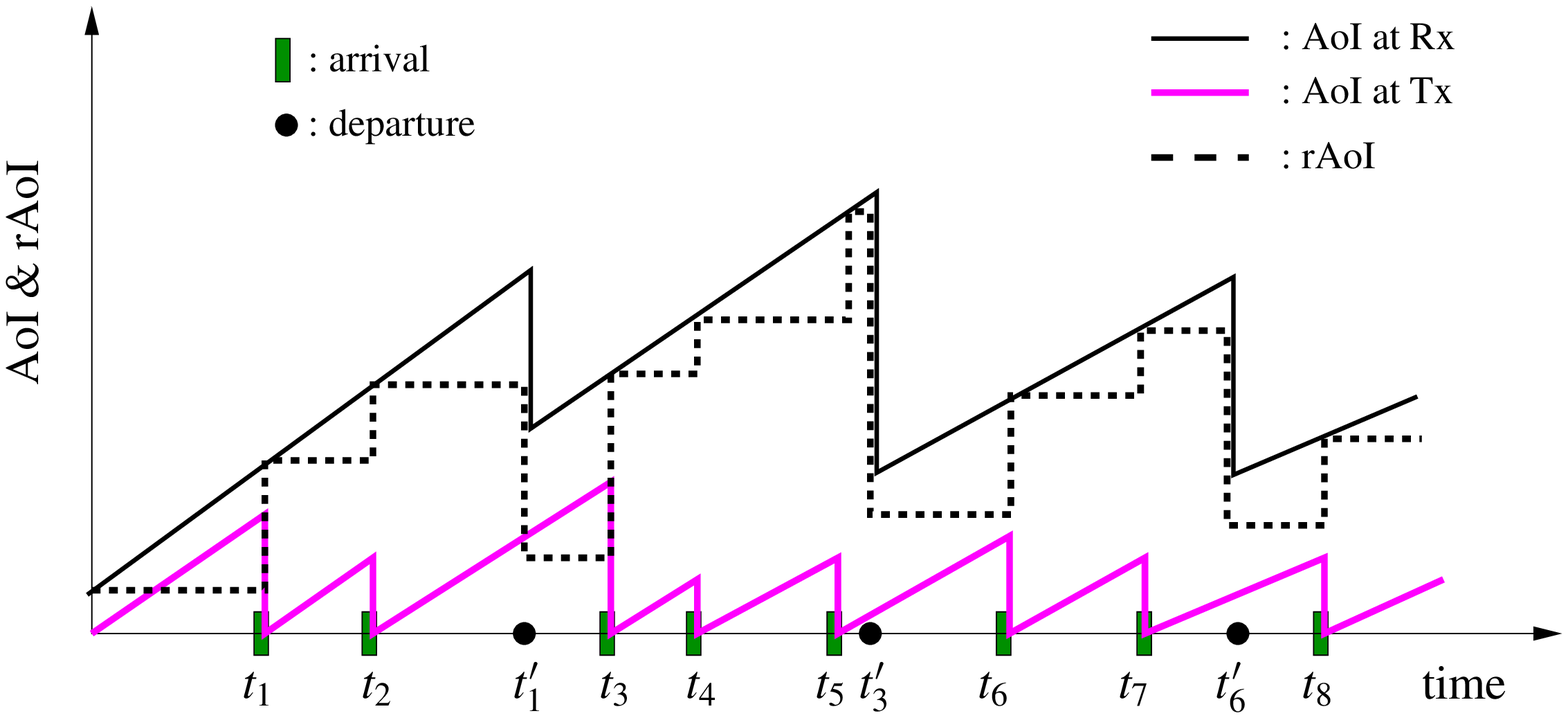}}
\caption{\sl Evolutions of $\Delta_R$, $\Delta_T$ and $\Gamma$ for M/GI/1/1.}\vspace{-0.2in}
\label{fig:4} 
\end{figure}

We next consider M/GI/1/1 scheme (see, e.g., \cite{najm2017status}) without preemption, where there is no buffer space for queuing and a packet can enter the server only if it is idle. We denote the two possible states of the system as (I) for idle and (B) for busy. Fig. \ref{fig:4} presents an instance of AoI and rAoI  evolutions under M/GI/1/1 packet management. Observe that at $t_2$, $t_4$, $t_5$ and $t_7$, the packets 2, 4, 5 and 7 are discarded upon their arrivals as they find the server in (B) state. $Q_i$ conditioned on the arriving packet finding the server in (I) is:
\begin{equation}
Q_i | (I) = \left \{  
\begin{array}{cc} (\Delta_R(t_i))^k X_i  &  \textrm{if $X_i < S_i$}   \\
(\Delta_R(t_i))^k S_i  &     \textrm{if $X_i \geq S_i $}
\end{array}
\right.
\label{eq:}
\end{equation}
Hence, $\mathbb{E}[Q | (I)]$ is identical to the expression in (\ref{eq:finexp}). Similarly, we have: 
\begin{equation}\hspace{-0.1in}
Q_i | (B) = \left \{  
\begin{array}{cc}
(\Delta_R(t_i))^k X_i  &  \textrm{if $X_i < \eta_i$}   \\
(\Delta_R(t_i))^k \eta_i + (S^{c} - \eta_i)^k(X_i - \eta_i)   & \textrm{if $X_i \geq \eta_i$}
\end{array}
\right.
\label{eq:}
\end{equation}
where $\eta_i$ represents the residual service time for the arriving packet. Note that conditioned on (I), $\eta_i=0$ and we use $\eta_i$ especially conditioned on (B). $S^{c}$ is the service time for the packet currently being served conditioned on the fact that it is greater than or equal to $\eta_i$. Here $S^c$ and $\eta_i$ are not independent; however, $\eta_i$ is independent of $X_i$. Residual time $\eta_i$ has the following density and moment generating functions (c.f. \cite[Eq. (36)]{inoue2018general}):
\begin{align}\label{kk1}
f_{\eta}(r)&=\frac{\mathbb{P}[S>r]}{\mathbb{E}[S]} \\
MGF^{(\eta)}_{\gamma} &= \frac{1-MGF^{(S)}_{\gamma} }{\gamma \mathbb{E}[S]} \label{kk2}
\end{align}
We evaluate $\mathbb{E}[Q_i|(B)]=\mathbb{E}[(\Delta_R(t_i))^k](\frac{1}{\lambda}-\frac{1}{\lambda}MGF_{\lambda}^{(\eta)}) + A$ where $A$ is the area due to the second term under condition $X_i \geq \eta_i$ that is calculated as
\begin{align*}
A=\frac{1}{\lambda}\int_{0}^{\infty}\int_{r}^{\infty}e^{-\lambda r} (s-r)^kf_S(s)f_{\eta}(r)dsdr
\end{align*}
Finally, the stationary probabilities of an arriving packet finding the server in (I) and (B) states are
\begin{align}
p_{I}=\frac{1}{1+\lambda \mathbb{E}[S]}, \ p_{B}=\frac{\lambda \mathbb{E}[S]}{1+ \lambda \mathbb{E}[S]}
\end{align} 
which follow due to PASTA property and the renewal structure already explored in \cite{najm2017status}. We can then calculate \[ \mathbb{E}[Q]=p_I \mathbb{E}[Q|(I)] + p_B \mathbb{E}[Q|(B)] \] and then $\mathbb{E}[\Gamma]=\lambda \mathbb{E}[Q]$.

\subsubsection{Exponential Service}
We now set $f_S(s)=\mu e^{-\mu s}$ for $s \geq 0$ and use the moment generating function expression in (\ref{eq:mgf_exp}). Note also that in this case $S$ and $\eta$ have identical distribution functions and $MGF_{\gamma}^{(S)}=MGF_{\gamma}^{(\eta)}$. Due to \cite[Eq. (21)]{costa2016age}, we have 
\begin{align}
\mathbb{E}[\Delta_R]=\frac{2\lambda^2+2\lambda\mu+\mu^2}{\lambda\mu(\lambda+\mu)}
\end{align}
Then, the first moment of rAoI is $\mathbb{E}[\Gamma]=\frac{2\lambda^2+\lambda\mu}{\lambda\mu(\lambda+\mu)}$. The second moment of AoI can be obtained using the technique in \cite[Theorem 1]{najm2017status} in the following form:
\begin{align}
\mathbb{E}[\Delta_R^2]=\frac{\lambda_e}{3}\mathbb{E}[(S_{i-1}+Y_{i-1})^3-S_i^3]
\end{align}  
where $Y=X+S$ marginally, $X$ is an inter-arrival time, $Y_i$ is independent of $S_i$ and $\lambda_e = \frac{1}{\mathbb{E}[X]+\mathbb{E}[S]}$. Working on the expression, we can then obtain
\begin{align}\nonumber
\mathbb{E}[\Delta_R^2]&=\frac{1}{3(\mathbb{E}[X]+\mathbb{E}[S])}\Bigg(\mathbb{E}[X^3]+\mathbb{E}[S^3]+6\mathbb{E}[X^2]\mathbb{E}[S]\\ &\ +6\mathbb{E}[X]\mathbb{E}[S^2]+6\mathbb{E}[X]\mathbb{E}^2[S]+6\mathbb{E}[S]\mathbb{E}[S^2]\Bigg)\label{eq:mg11}
\end{align}
In the expression in (\ref{eq:mg11}), we plug $\mathbb{E}[X]=\frac{1}{\lambda}$, $\mathbb{E}[X^2]=\frac{2}{\lambda^2}$, $\mathbb{E}[X^3]=\frac{6}{\lambda^3}$ and similarly $\mathbb{E}[S]=\frac{1}{\mu}$, $\mathbb{E}[S^2]=\frac{2}{\mu^2}$, $\mathbb{E}[S^3]=\frac{6}{\mu^3}$. Next, we have $\mathbb{E}[Q|(I)]=\mathbb{E}[\Delta_R^2]\frac{1}{\lambda+\mu}$, $\mathbb{E}[Q|(B)]=\mathbb{E}[\Delta_R^2]\frac{1}{\lambda+\mu}+A$, $p_I=\frac{\mu}{\lambda+\mu}$ and $p_B=\frac{\lambda}{\lambda+\mu}$. It remains to determine $A$:
\begin{align*}
A=\frac{1}{\lambda}\int_{0}^{\infty}\int_{0}^{\infty} e^{-(\lambda+\mu) r} s^2f_S(s)f_{\eta}(r)dsdr = \frac{2\mu}{\lambda^3(\lambda + 2\mu)}
\end{align*}
We then combine the terms to obtain the following:
\begin{align}
\mathbb{E}[\Gamma^2]=(\mathbb{E}[\Delta_R^2]+\lambda A)\frac{\lambda}{\lambda+\mu}
\end{align}

\subsubsection{Deterministic Service}
In this case, we set $S=\frac{1}{\mu}$ and use the moment generating function expressions in (\ref{eq:mgf_det}). Due to \cite{najm2017status}, we have 
\begin{align}
\mathbb{E}[\Delta_R]=\frac{3\lambda^2+4\lambda\mu+2\mu^2}{2\lambda\mu(\lambda+\mu)}
\end{align}
Then, we have first moment of rAoI as $\mathbb{E}[\Gamma]=\frac{3\lambda^2+2\lambda\mu}{2\lambda\mu(\lambda+\mu)}$. To get the second moment, we use equation (\ref{eq:mg11}) and set $\mathbb{E}[S]=\frac{1}{\mu}$, $\mathbb{E}[S^2]=\frac{1}{\mu^2}$ and $\mathbb{E}[S^3]=\frac{1}{\mu^3}$ with the same statistics for $X$. Next, we have $\mathbb{E}[Q|(I)]=\mathbb{E}[\Delta_R^2]\frac{1}{\lambda}(1-e^{-\frac{\lambda}{\mu}})$, $p_I=\frac{\mu}{\lambda+\mu}$ and $p_B=\frac{\lambda}{\lambda+\mu}$. It remains to determine $A$:
\begin{align*}
A&= \int_{0}^{\infty}\int_{r}^{\infty}\frac{e^{-\lambda r}}{\lambda} (s-r)^kf_S(s)f_{\eta}(r)dsdr\\
&=\mu\int_{0}^{\frac{1}{\mu}} \frac{e^{-\lambda r}}{\lambda} (\frac{1}{\mu}-r)^2dr =\frac{\mu e^{-\frac{\lambda}{\mu}}}{\lambda}\int_0^{\frac{1}{\mu}} u^2 e^{\lambda u}du \\ &=\frac{\mu}{\lambda^4}\left(\frac{\lambda^2}{\mu^2}-2\frac{\lambda}{\mu}+2 - 2e^{-\frac{\lambda}{\mu}}\right)
\end{align*}
We then get $\mathbb{E}[Q|(B)]=\mathbb{E}[\Delta_R^2]\frac{1}{\lambda}(1-\frac{\mu\left(1-e^{-\frac{\lambda}{\mu}}\right)}{\lambda})  +A$. We finally combine the terms to obtain the following:
\begin{align}\nonumber
\mathbb{E}[\Gamma^2]&=\mathbb{E}[\Delta_R^2]\left((1-e^{-\frac{\lambda}{\mu}})p_I + (1-\frac{\mu\left(1-e^{-\frac{\lambda}{\mu}}\right)}{\lambda})p_B\right) \\ &\qquad+\lambda Ap_B
\end{align}

\subsection{M/GI/1/$2^*$}

\begin{figure}[!t]
\centering{
\hspace{-0.2cm} 
\includegraphics[totalheight=0.19\textheight]{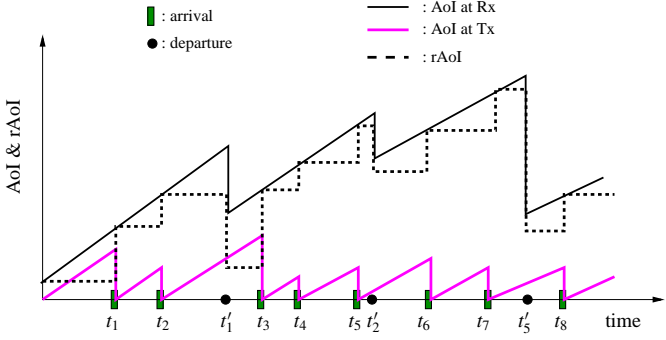}}
\caption{\sl Evolutions of $\Delta_R$, $\Delta_T$ and $\Gamma$ for M/GI/1/2*.}\vspace{-0.2in}
\label{fig:5} 
\end{figure}

We finally consider M/GI/1/$2^*$ scheme (see M/M/1/$2^*$ in \cite{costa2016age}) or equivalently non-preemptive last come first serve with discarding (see \cite{inoue2018general}). In this scheme, we assume that a single space buffer is available for queuing. When the server is busy, the transmitter keeps the latest arriving update in the buffer and discards the previous updates. We provide an illustration for this scheme in Fig. \ref{fig:5}. Assuming (I) initial state, packet 1 enters the server right away and packet 2 is kept in the queue until $t_1'$ when it is taken to service. Then, packet 1's service ends, packet 3 arrives and is kept in the buffer until $t_4$ when it is replaced with arriving packet 4. Then, packet 4 is discarded when packet 5 arrives at $t_5$. Finally, during packet 5's service, packet 6 arrives first and then it is replaced with the newer arrival packet 7.

With Poisson arrivals, general independent service time distribution, a single server, and a single space in the buffer, this is in the form of an M/GI/1/$2^*$ queue in Kendall notation. We again condition on two states of the server (I) and (B). We denote the remaining time until the end of service at the instant of packet $i$'s arrival as $\zeta_i$. We also denote the service time for the packet that enters service at the end of $\zeta_i$ as $S_i$, an independent random variable with density $f_S(s)$.  
\begin{equation}
Q_i | (I) = \left \{  
\begin{array}{cc} (\Delta_R(t_i))^k X_i  &  \textrm{if $X_i < S_i$}   \\
(\Delta_R(t_i))^k S_i  &     \textrm{if $X_i \geq S_i $}
\end{array}
\right.
\label{eq:}
\end{equation}
and therefore $\mathbb{E}[Q | (I)]$ is identical to the expression in (\ref{eq:finexp}). Now, we have: 
\begin{equation}\hspace{-0.1in}
Q_i | (B) = \left \{  
\begin{array}{cc}
(\Delta_R(t_i))^k X_i  &  \textrm{if $X_i < \zeta_i$}   \\
(\Delta_R(t_i))^k \zeta_i + \\ (T^{c} - \zeta_i)^k(X_i - \zeta_i)   & \textrm{if $\zeta_i + S_i \geq X_i \geq \zeta_i$} \\
(\Delta_R(t_i))^k \zeta_i + \\ (T^{c} - \zeta_i)^kS_i   & \textrm{if $\zeta_i + S_i < X_i$}
\end{array}
\right.
\label{eq:}
\end{equation}
where $\zeta_i$ represents the residual service time for the arriving packet and $T^{c}$ is the system time for the packet in service conditioned on the fact that it is greater than or equal to $\zeta_i$. Observe that $\Gamma(t)$ drops to zero in the interval $[t_i+\zeta_i + S_i, t_i+X_i]$ if $\zeta_i + S_i < X_i$. Here $T^c$ and $\zeta_i$ are not independent; still, $X_i$, $\zeta_i$ and $S_i$ are mutually independent. $\zeta_i$ has the same probability density and moment generating functions as in (\ref{kk1})-(\ref{kk2}). \[ \mathbb{E}[Q_i|(B)]=\mathbb{E}[(\Delta_R(t_i))^k](\frac{1}{\lambda}-\frac{1}{\lambda}MGF_{\lambda}^{(\zeta)}) + K \] where $MGF_{\lambda}^{(\zeta)}$ is the moment generating function for $\zeta$ and $K$ is the combined area due to the second terms under conditions $\zeta_i + S_i \geq X_i \geq \zeta_i$ and $\zeta_i + S_i < X_i$. We have
\begin{align*}
K&=\frac{(1-MGF_{\lambda}^{(S)})}{\lambda}\int_{0}^{\infty}\int_{r}^{\infty}e^{-\lambda r}(t-r)^kf_T(t)  f_{\zeta}(r)dtdr 
\end{align*}
The probability density and moment generating functions for system time $T$ are calculated in terms of those of service distribution following the steps in \cite[Appendix E.2]{inoue2018general}.
Finally, the stationary probabilities for an arriving packet finding the system in (I) and (B) states are obtained as follows (see \cite{infocom_arxiv}):
\begin{align}\label{one}
p_{I} = \frac{MGF^{(S)}_{\lambda}}{MGF^{(S)}_{\lambda}+\lambda \mathbb{E}[S]}, \ p_{B}= \frac{\lambda \mathbb{E}[S]}{MGF^{(S)}_{\lambda}+\lambda \mathbb{E}[S]}
\end{align}
We calculate $\mathbb{E}[Q]=p_I \mathbb{E}[Q|(I)] + p_B \mathbb{E}[Q|(B)]$ and then $\mathbb{E}[\Gamma]=\lambda \mathbb{E}[Q]$.

\subsubsection{Exponential Service}
We now consider $f_S(s)=\mu e^{-\mu s}$ for $s \geq 0$ and refer to the moment generating function expressions in (\ref{eq:mgf_exp}). Due to \cite[Eq. (65)]{costa2016age}, we have 
\begin{align*}
\mathbb{E}[\Delta_R]=\frac{1}{\lambda}+\frac{2}{\mu}+\frac{\lambda}{(\lambda+\mu)^2}+\frac{1}{\lambda+\mu}-\frac{2(\lambda+\mu)}{\lambda^2+\lambda\mu+\mu^2}
\end{align*}
The first moment of rAoI is $\mathbb{E}[\Gamma]=\frac{2}{\mu}+\frac{\lambda}{(\lambda+\mu)^2}+\frac{1}{\lambda+\mu}-\frac{2(\lambda+\mu)}{\lambda^2+\lambda\mu+\mu^2}$. We can get the second moment of AoI, $\mathbb{E}[\Delta^2_R]$, through the second derivative of Laplace-Stieltjes transform of the age in \cite[Section A.3]{inoue2018general} and evaluating at $s=0$. We use MATLAB symbolic tool to evaluate the second derivative. The resulting expression is in closed form and plotted easily with MATLAB.  To obtain $\mathbb{E}[Q]$, it remains to determine the variable $K$:
\begin{align*}
K=\frac{1}{\lambda+\mu}\int_{0}^{\infty}\int_{r}^{\infty}e^{-\lambda r}(t-r)^2f_T(t)  f_{\zeta}(r)dtdr 
\end{align*}
where $f_{\zeta}(r)$ is the density function of residual service time for a packet and it is identical to $f_S(s)$ due to memoryless service. We take the derivative of \cite[Eq. (59)]{costa2016age} to obtain
\begin{align}
f_T(t)=c_1e^{-\mu t} - c_2 e^{-(\lambda + \mu)t}
\end{align}
where $c_1=\frac{\mu p_I}{p_I + \frac{\mu}{\lambda+\mu}(1-p_I)}(1+\frac{\mu}{\lambda})$ and $c_2=\frac{\frac{\mu^3}{\lambda} (1-p_I)}{p_I + \frac{\mu}{\lambda+\mu}(1-p_I)}$. Here, $p_I=\frac{\mu^2}{\lambda^2+\lambda\mu+\mu^2}$ is the idle probability from equation (\ref{one}). We then have $K=K_1-K_2$ where
\begin{align}
K_1&=\frac{c_1}{\lambda+\mu}\int_{0}^{\infty}\int_{0}^{\infty}e^{-(\lambda+\mu) r}t^2e^{-\mu t}  f_{\zeta}(r)dtdr \\
K_2&=\frac{c_2}{\lambda+\mu}\int_{0}^{\infty}\int_{0}^{\infty}e^{-(2\lambda+\mu) r}t^2e^{-(\lambda+\mu)t}  f_{\zeta}(r)dtdr 
\end{align}
Then, we calculate $K_1=\frac{2c_1}{(\lambda+\mu)\mu^2(\lambda+2\mu)}$ and $K_2=\frac{c_2\mu}{(\lambda+\mu)(\lambda+\mu)^3(\lambda+\mu)}$.
We then combine the terms to obtain:
\begin{align}
\mathbb{E}[\Gamma^2]=(\mathbb{E}[\Delta_R^2]+\lambda K)\frac{\lambda}{\lambda+\mu}
\end{align}\vspace{-0.2in}

\subsubsection{Deterministic Service}
In this case, we consider $S=\frac{1}{\mu}$ and refer to the moment generating function expressions in (\ref{eq:mgf_det}). Due to \cite[Section A.2]{inoue2018general}, we have 
\begin{align}
\mathbb{E}[\Delta_R]=\frac{1}{\mu}\left(\frac{3}{2} + \frac{\mu e^{\frac{\lambda}{\mu}} - \lambda - \mu}{\lambda e^{\frac{\lambda}{\mu}}} + \frac{(\lambda+2\mu)\mu}{2\lambda (\mu + \lambda e^{\frac{\lambda}{\mu}}) }\right)
\end{align}
Then, we get the first moment of rAoI as $\mathbb{E}[\Gamma]=\mathbb{E}[\Delta_R]-\frac{1}{\lambda}$. 
The second moment of AoI, $\mathbb{E}[\Delta^2_R]$, can be obtained by the second derivative of Laplace-Stieltjes transform of the age in \cite[Theorem 47]{inoue2018general} and evaluating at $s=0$. We again use MATLAB symbolic tool to evaluate the second derivative. To obtain $\mathbb{E}[Q]$, we will determine the variable $K$. To this end, we need the density function $f_T(t)$. From the transfrom in \cite[Lemma 46, Eq. (63)]{inoue2018general}, we can obtain
\begin{align}\nonumber
f_T(t)=e^{-\frac{\lambda}{\mu}}\delta(t-\frac{1}{\mu}) + \lambda e^{- \lambda (t- \frac{1}{m})}(u(t-\frac{1}{\mu}) - u(t-\frac{2}{\mu}))
\end{align}
where $\delta(t)$ is Dirac delta function and $u(t)$ is the unit step function. We then have $K=K_1+K_2$ where
\begin{align*}
K_1&=\frac{(1-e^{-\frac{\lambda}{\mu}})e^{-\frac{\lambda}{\mu}}\mu}{\lambda}\int_{0}^{\frac{1}{\mu}}e^{-\lambda r}(\frac{1}{\mu}-r)^2 dr \end{align*} 

\begin{align*}
K_2&=\frac{(1-e^{-\frac{\lambda}{\mu}})\mu}{\lambda}\int_{0}^{\frac{1}{\mu}}e^{-\lambda r}\int_{\frac{1}{\mu}}^{\frac{2}{\mu}}(t-r)^2e^{-\lambda(t-\frac{1}{\mu})} dtdr 
\end{align*}
Then, $K_1=\frac{(1-e^{-\frac{\lambda}{\mu}})e^{-\frac{\lambda}{\mu}}\mu}{\lambda^4}\left(\frac{\lambda^2}{\mu^2}-2\frac{\lambda}{\mu}+2 - 2e^{-\frac{\lambda}{\mu}}\right)$ and $K_2=\frac{(1-e^{-\frac{\lambda}{\mu}})\mu}{\lambda^5}(2+\frac{\lambda^2}{\mu^2}-e^{-\frac{\lambda}{\mu}}(2+4\frac{\lambda^2}{\mu^2})+e^{-2\frac{\lambda}{\mu}}(\frac{\lambda^2}{\mu^2}+2))$. We then combine the terms as follows:
\begin{align}\nonumber
\mathbb{E}[\Gamma^2]&=\mathbb{E}[\Delta_R^2]\left((1-e^{-\frac{\lambda}{\mu}})p_I + (1-\frac{\mu\left(1-e^{-\frac{\lambda}{\mu}}\right)}{\lambda})p_B\right) \\ &\qquad+\lambda K p_B
\end{align}
where $p_B=\frac{\lambda}{\lambda +\mu e^{-\frac{\lambda}{\mu}}}$ and $p_I=1-p_B$ from equation (\ref{one}).

\section{Numerical Results}
\label{sec:Numres}

In this section, we provide numerical results for the first and second moments of rAoI with respect to system parameters under exponential and deterministic service distributions. Note that in view of Remark \ref{re:1} the variation of the first moment with respect to the service rate for fixed arrival rate is identical to that previously reported for AoI, except for an additional shift. Therefore, we pay special attention to fixed service rate and rAoI as a function of $\lambda$. In all numerical results, we performed packet-based queue simulations for $10^6$ packets as verification and each time we observed the plots are compatible.

\begin{figure}[!t]
\centering{
\hspace{-0.6cm} 
\includegraphics[totalheight=0.30\textheight]{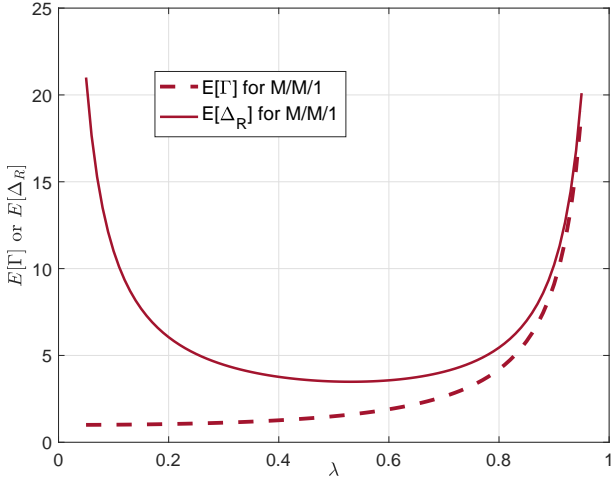}}\vspace{-0.2in}
\caption{\sl First moments of AoI and rAoI versus arrival rate $\lambda$ for fixed $\mu=1$ under M/M/1. }\vspace{-0.2in}
\label{fig:num4} 
\end{figure}

We start with Fig. \ref{fig:num4} where we compare the first moments of AoI and rAoI under M/M/1 with no packet management for fixed $\mu=1$. We use \cite[Eq. (17)]{kaul2012real} to calculate $\mathbb{E}[\Delta_R]$ and then use Remark \ref{re:1} to get $\mathbb{E}[\Gamma]$. The classical AoI is large for small $\lambda$ because updates age significantly when they are not generated frequently. In contrast, rAoI is small for small $\lambda$ (with a minimum at $\lambda = 0$) because it is easier for the receiver to synchronize with the update at the transmitter in this case and the age is essentially equal to the packet delay. As $\lambda$ increases, rAoI monotonically increases and approaches the classical AoI and $\infty$ as queuing delays mount up.

\begin{figure}[!t]
\centering{
\hspace{-0.6cm} 
\includegraphics[totalheight=0.30\textheight]{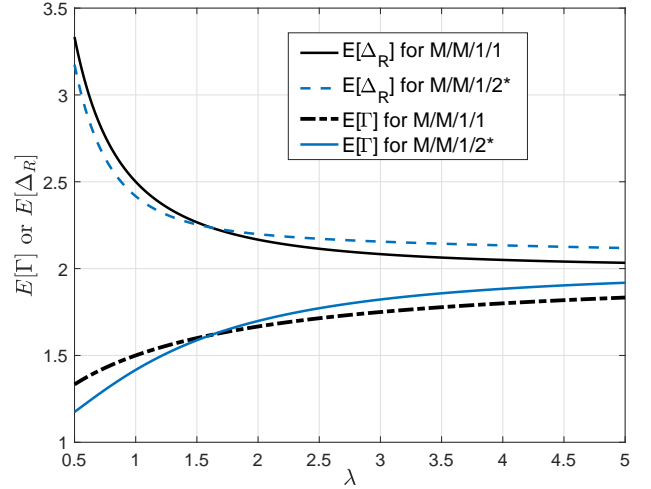}}\vspace{-0.2in}
\caption{\sl The comparison of the first moments of AoI and rAoI with respect to arrival rate $\lambda$ for fixed $\mu=1$. }\vspace{-0.1in}
\label{fig:num3} 
\end{figure}

We next move on to systems with limited buffering and packet management. We show in Fig. \ref{fig:num3} the comparison among the first moments of AoI and rAoI with respect to $\lambda$ for both M/M/1/1 and M/M/1/$2^*$. We observe that as $\lambda$ grows to infinity, both rAoI and AoI converge as predicted analytically in Remark \ref{re:1}. It is interesting to note that the AoI monotonically decreases with $\lambda$ whereas rAoI monotonically increases. For small $\lambda$, updates are delivered with small delays and as the rAoI measures the timeliness of delivering the updates to the receiver after they are generated, it remains small. The AoI, on the other hand, is large because it measures the time since the last update at the receiver. This captures the essential difference between the two metrics. The AoI captures the absolute age of status updates without considering the update generation frequency that may be appropriate for a particular source. The rAoI captures the efficiency of the update delivery system in isolation\footnote{We note that this may appear similar to packet delay, but delay for a given packet is not affected by later arrivals, whereas rAoI is updated for every arrival event.}. We also note the subtle difference between the two queuing systems. For small $\lambda$, both AoI and rAoI are smaller for M/M/1/$2^*$ than for M/M/1/1 as the additional buffer space helps in improving the age of highly infrequent updates. As the arrival rate increases, M/M/1/1 becomes better by virtue of its lower system time for each packet served.

\begin{figure}[!t]
\centering{
\hspace{-0.6cm} 
\includegraphics[totalheight=0.30\textheight]{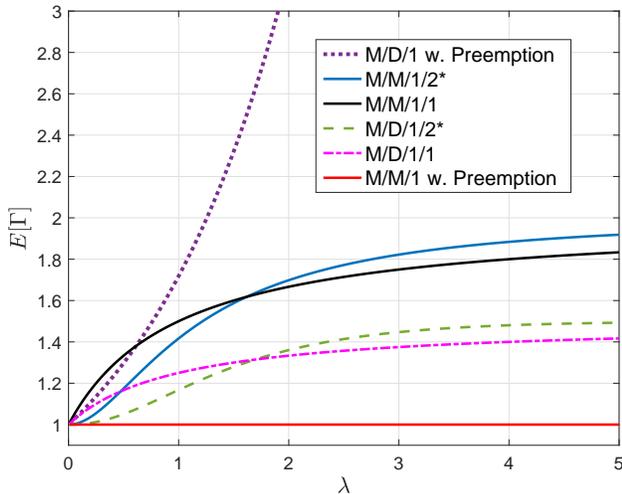}}\vspace{-0.2in}
\caption{\sl This plot shows the variation of the first moment of rAoI with respect to arrival rate $\lambda$ for fixed $\mu=1$.}\vspace{-0.2in}
\label{fig:num1} 
\end{figure}

We focus exclusively on rAoI from here on. In Fig. \ref{fig:num1}, we plot the first moment of rAoI with varying $\lambda$ for fixed $\mu=1$ for all the packet management and service time distributions we explored in this paper. We observe that the presence of preemption yields different outcomes in terms of $\mathbb{E}[\Gamma]$ for different service distributions. On the one hand, for memoryless exponential distribution, we have shown that $\mathbb{E}[\Gamma] = 1/\mu$ and is therefore invariant with respect to $\lambda$,  whereas $\mathbb{E}[\Gamma]$ diverges to infinity very quickly as $\lambda$ goes beyond unity under deterministic service distribution. Note that for deterministic service distribution, as $\lambda$ increases the likelihood of a new arrival during a service interval increases and this causes the server to never be able to finish service. This issue is not observed for memoryless service. We also observe that as $\lambda \rightarrow \infty$ the limiting $\mathbb{E}[\Gamma]$ are both equal to 2 for M/M/1/1 and M/M/1/$2^*$ while the limit is equal to $\frac{3}{2}$ for both M/D/1/1 and M/D/1/$2^*$. In the comparison between M/./1/1 and M/./1/$2^*$, the latter one has better mean rAoI performance for smaller loads while the opposite is true for larger loads. It is also remarkable that M/M/1 with preemption leads to the smallest rAoI uniformly.

\begin{figure}[!t]
\centering{
\hspace{-0.6cm} 
\includegraphics[totalheight=0.30\textheight]{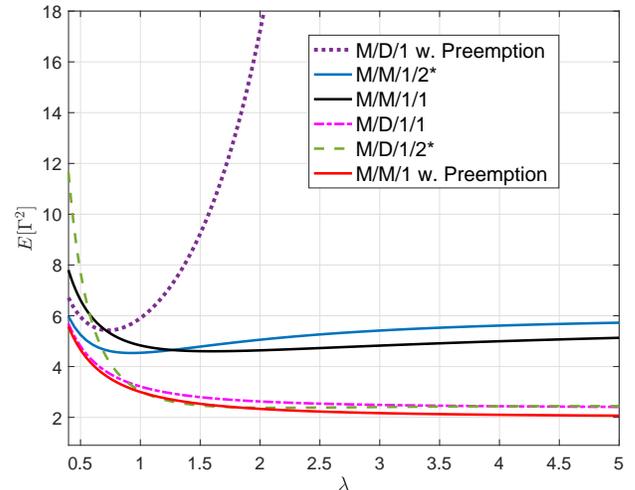}}\vspace{-0.2in}
\caption{\sl The variation of the second moment of rAoI with respect to arrival rate $\lambda$ for fixed $\mu=1$. }\vspace{-0.2in}
\label{fig:num2} 
\end{figure}

In Fig. \ref{fig:num2}, we present the second moment of rAoI with respect to $\lambda$ for fixed $\mu=1$. We observe very similar trends to those we observed for the first moment of rAoI in Fig. \ref{fig:num1}. In particular, the second moment takes uniformly the smallest value under M/M/1 with preemption whereas the M/D/1 with preemption case yields a diverging second moment for rAoI. Additionally, as $\lambda$ grows to infinity the point of convergence for M/M/1/1 (or M/D/1/1) is identical to M/M/1/2* (or M/D/1/2*).

\section{Conclusions}
\label{sec:Conc}

In this paper, we introduce relative Age of Information (rAoI) metric and analyze it for various packet management schemes. This new metric aims to capture cases of undetermined timings of data generation at the source as is typically the case for decentralized applications. In such cases, transmission schedules are blind to the data generation timing and \textit{an update packet remains fresh until a new one arrives.} The rAoI metric measures how fresh the data is at the receiver \textit{relative to} the transmitter. We provide closed form expressions to calculate moments of rAoI from the moments of classical AoI, applicable to a wide range of service distributions with memoryless arrivals. In particular, we address M/GI/1 with preemption, M/GI/1/1 and M/GI/1/$2^*$ cases. In the numerical results, we focus on memoryless exponential and deterministic service distributions. Our numerical results reveal several interesting behaviors of the first and second moments of rAoI with varying system load.

\end{document}